\documentclass[%
 reprint,
superscriptaddress,
 amsmath,amssymb,
 aps,prl
]{revtex4-1}

\usepackage{graphicx}% Include figure files
\usepackage{dcolumn}% Align table columns on decimal point
\usepackage{bm}% bold math
\usepackage{soul,color}
\usepackage{float}
\begin{document}

\preprint{APS/123-QED}

%\title{Towards 1\% single photon nonlinearity \\ in a lithium niobate microring resonator with a second-harmonic generation efficiency of 5,000,000\%/W}

\title{Towards 1\% single photon nonlinearity \\ with periodically-poled lithium niobate microring resonators }

%\title{Prospects for single-photon nonlinearity \\ in periodically-poled lithium niobate microring resonators \\ with a second-harmonic generation efficiency of 5,000,000\%/W}% Force line breaks with \\

\author{Juanjuan Lu}
\affiliation{Department of Electrical Engineering, Yale University, New Haven, Connecticut 06511, USA}%
\author{Ming Li}
\author{Chang-Ling Zou}
\affiliation{Department of Optics, University of Science and Technology of China, Hefei 230026, P. R. China.}
\author{Ayed Al Sayem}
\author{Hong X. Tang}
 \email{hong.tang@yale.edu}
\affiliation{Department of Electrical Engineering, Yale University, New Haven, Connecticut 06511, USA}

\date{\today}% It is always \today, today,

\begin{abstract}
The absence of the single-photon nonlinearity has been a major roadblock in developing quantum photonic circuits at optical frequencies. In this paper, we demonstrate a periodically-poled thin film lithium niobate microring resonator (PPLNMR) that reaches 5,000,000\%/W second harmonic conversion efficiency---almost 20-fold enhancement over the state-of-the-art---by accessing its largest $\chi^{(2)}$ tensor component $d_{33}$ via quasi-phase matching. The corresponding single photon coupling rate $g/2\pi$ is estimated to be 1.2~MHz, which is an important milestone as it approaches the dissipation rate $\kappa/2\pi$ of best available lithium niobate microresonators developed in the community. Using a figure of merit defined as $g/\kappa$, our devices reach a single photon nonlinearity approaching 1\%. We show that, by further scaling of the device, it is possible to improve the single photon nonlinearity to a regime where photon-blockade effect can be manifested.  
\end{abstract}

\maketitle

%\tableofcontents

\section{\label{sec:intro}Introduction}
Quantum photonic integrated circuits have received growing interests since such platforms offer the stability and integrability towards solid state quantum applications~\cite{OBrien2009,Politi1221,Aspuru12,doi:10.1063/1.5100160,Wang2019,Elshaari}. By encoding quantum information into the optical photons, the quantum information processing, quantum communication and quantum metrology would benefit from the merits of the bosonic carriers, including the high propagation velocity, long propagation distance and infinite-dimensional Hilbert spaces. Microwave photons could be processed by superconducting quantum circuits, where high fidelity quantum operations approaching error-correction thresholds are achieved by the lossless nonlinearity inherent to Josephson effect~\cite{Girvin2014, wendin2017quantum}. However, at optical frequencies, the absence of the single-photon nonlinearity hinders the quest for the deterministic single-photon sources and further high-fidelity photonic quantum gates, which are two main building blocks for the scalable quantum circuits~\cite{Kok2007,Politi646,crespi2011,Khasminskaya2016,caspani2017}. While the photon–photon interactions required to realize quantum gate can be mediated through light–matter interaction with atomic or atom-like solid-state emitters, routes to scaling such systems remain challenging~\cite{Aharonovich2016,Awschalom2018,lukin14}.

Leveraging the small mode volume and enhanced photon-photon interaction in high quality-factor (Q) optical microcavities~\cite{Vahala2003} as well as the inherent material nonlinearity~\cite{boyd2019nonlinear}, several schemes based on the microcavity made by the materials with second-order ($\chi^{(2)}$) nonlinearity have been theoretically proposed to allow the indistinguishable single-photon state, as well as the deterministic photon-photon gates via photon blockade effect by harnessing the single-photon nonlinearity~\cite{PhysRevLett.96.057405,Majumdar2013,Heuck2020,Ming2020}, and thereby bring in-sight scalable quantum photonic computing. Lithium niobate (LN) has recently risen to the forefront of integrated quantum photonics circuits due to its intrinsic strong optical nonlinearity, electro-optic effect and experimentally demonstrated ultralow-loss nanophotonics platform for scaling. In this letter, we describe a 20-fold enhancement over state-of-the-art devices in second harmonic generation (SHG)~\cite{Lu2019d,Chen:19} with the thin film periodically poled lithium niobate microring resonators (PPLNMRs) by leveraging its largest $\chi^{(2)}$ tensor element $d_{33}$ for quasi-phase matching. This exceptionally high nonlinearity translates to a vacuum photon-photon coupling strength $g/2\pi$ of 1.2~MHz, which is an important milestone as it approaches the dissipation rate $\kappa$ of best available LN microresonators~\cite{zhang2017monolithic}. Currently, the trade-off between mode confinement and optical loss limits our device to a figure of merit (FOM) single photon nonlinearity, defined as $g/\kappa$, at $10^{-2}$. Upon further scaling of microresonators, it is possible to improve this FOM by an order of magnitude, making it feasible to realize the photon-blockade effect in certain specifically-designed device configuration. 

\begin{figure}[htbp]
\includegraphics[width=0.48\textwidth]{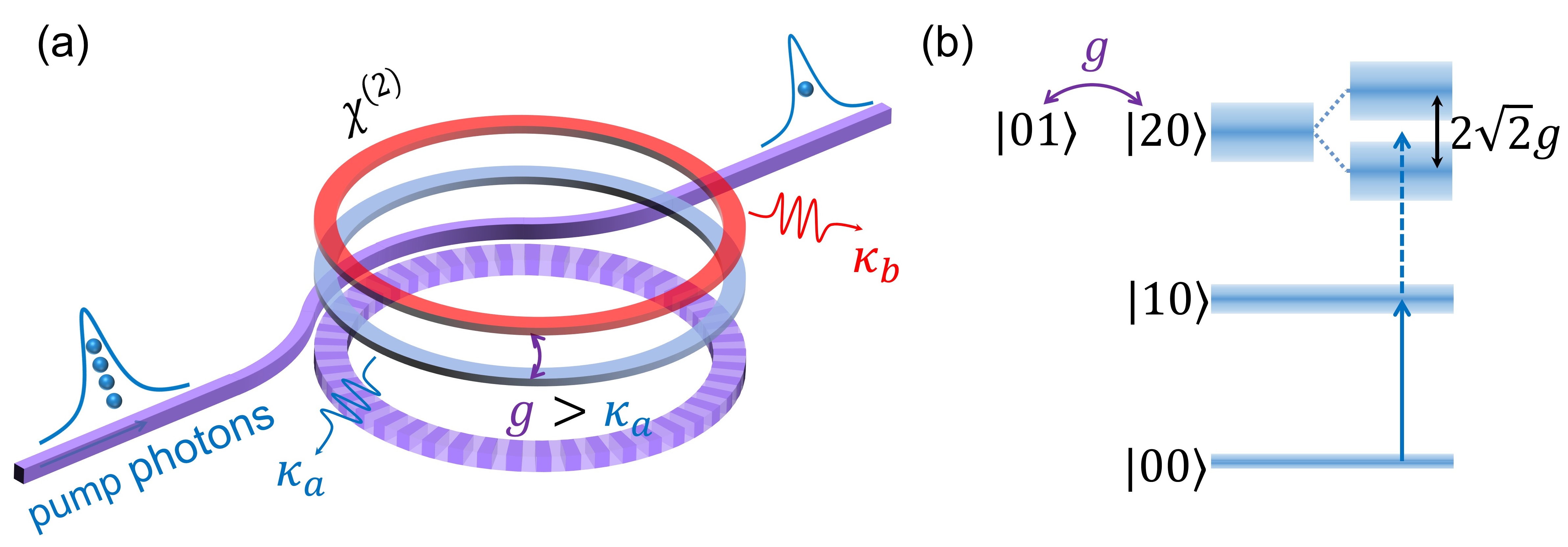}
\caption{\label{fig1} Photon blockade effect due to $\chi^{(2)}$ single photon nonlinearity. (a) A doubly-resonant cavity based on PPLNMR with single photon nonlinearity and thereby showing photon blockade effect. (b) The schematic energy level diagram of a PPLNMR with single photon nonlinearity. The anharmonicity of the nonlinear system is determined by the energy level splitting relative to their widths.}
\end{figure}  
Single photon nonlinearity is appealing for quantum photonics applications. As an example, Fig.~\ref{fig1} illustrates a conceptual photon-blockade device leveraging single-photon-level $\chi^{(2)}$ nonlinearity based on a PPLNMR, which could generate single photons with sub-poissonian quantum statistics from a classical laser input. Photons with carrier frequency $\omega_{a}$ (blue) travel in a waveguide and couple to the cavity mode $a$ with a dissipation rate $\kappa_{a}$. With the fulfillment of certain phase-matching condition for the degenerate three-wave mixing, a significant nonlinear coupling strength $g$ between mode $a$ and mode $b$ (with a frequency of $\omega_b=2\omega_a$) could be obtained and thereby a frequency conversion from $\omega_a$ to $\omega_b$ is feasible. For example, the highly efficient second-harmonic generation between mode $a$ in the telecom band and mode $b$ in the near-visible band has been experimentally demonstrated in a PPLNMR~\cite{Lu2019d}. The system Hamiltonian of such PPLNMR device reads
\begin{eqnarray}
\hat{H}/\hbar = &&\omega_{a}{\hat{a}}^{\dagger}\hat{a} + \omega_b\hat{b}^{\dagger}\hat{b}  + g((\hat{a}^{\dagger})^2\hat{b} + \hat{a}^{2}\hat{b}^{\dagger}).
\end{eqnarray}
Here, $\hat{a}$ and $\hat{b}$ represent the bosonic operators, $g$ denotes the vacuum nonlinear photon-photon interaction strength between mode $a$ and $b$. Considering only a few excitations, the system energy levels can be written as $|mn\rangle=|m\rangle_{a}\otimes|n\rangle_{b}$ in the Fock-state basis, with $m,n \in \mathbb{Z}$. When the single photon nonlinearity is realized, i.e. the coupling strength exceeds the dissipation rate, the state $|20\rangle$ strongly couples to $|01\rangle$, and new eigenstates $(|20\rangle\pm|01\rangle)/\sqrt{2}$ with a frequency splitting of $2\sqrt{2}g$ is produced [Fig.~\ref{fig1}(b)]. The induced anharmonicity of energy levels gives rise to the photon blockade effect as illustrated in Fig.~\ref{fig1}(a), where the pump mode energy level with photon number $N\geq2$ are no longer resonant with the cavity, and thereby the two-excitation state would be inhibited if the detuning $\sqrt{2}g$ is comparable or larger than the energy linewidth $\kappa_a$, as indicated by the blue dashed arrow in Fig.~\ref{fig1}(b). Hence, we introduce the dimensionless figure of merit quantifying the cooperation of nonlinear optical processes at the single-photon level: $\mathrm{FOM} = g/\kappa_a$. The larger FOM, the better performance in single photon generation. Alternatively, the FOM also indicates the number of quantum gate operations on single photons before significant fidelity loss due to the photon dissipation to environment.

\section{Optimization of coupling strength \lowercase{\boldmath{$g$}}}

Based on the above discussion, a larger coupling strength $g$ is always demanding to realize single-photon nonlinearity, since the mode dissipation rates are mostly restricted by the material and fabrication technique in practices. In a microring resonator, the photon-photon coupling strength $g$ is determined by the material $\chi^{(2)}$ coefficient, modal overlap factor $\gamma$, and the mode volume $V$ through the relation: $g\propto \chi^{(2)}\gamma / \sqrt{V}$. According to  Ref.~\cite{Lu2019d}, the PPLNMR is employed for high $\chi^{(2)}$ coefficient and phase matching condition, and a SHG efficiency of $250,000\%$ was achieved. Here, a significant improvement of $g$ to its theoretical limit is demonstrated by solving two practical challenges: the utilization of largest nonlinear term $d_{33}$ of the z-cut LN thin film and the high-fidelity poling of the microring. 

\begin{figure*}
\includegraphics[width=1.0\textwidth]{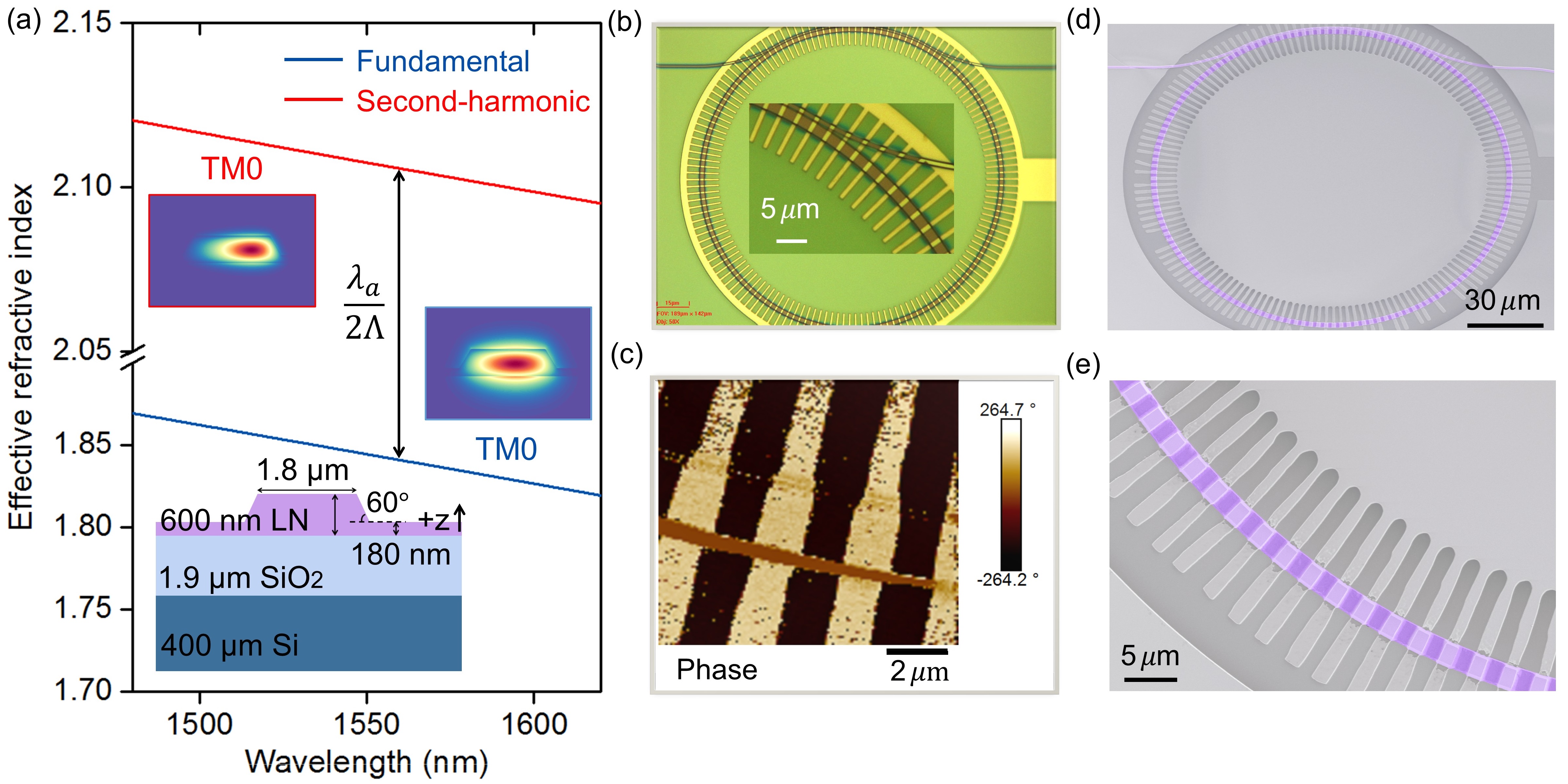}  
\caption{(a) Effective refractive indices of the TM fundamental and SH modes, where the criterion for the type-0, first-order QPM between mode $a$ at 1560~nm and mode $b$ at 780~nm with a poling period $\Lambda=2.95$~$\mu$m is indicated by the vertical, double-headed arrow. The upper insets show the simulated profiles (amplitude of the vertical electric-field component) of the mode $a$ and $b$ while The lower inset depicts the schematic cross section of a z-cut PPLNMR with a radius of 70~$\mu$m. (b) Optical image of the etched microring with radial poling electrodes and its zoomed view around the waveguide-microring coupling region. (c) PFM phase scan over a small portion of a PPLNMR, revealing the alternate ferroelectric domain structures and a duty cycle close to 50\%. (d) False-color SEM image of a deplicate PPLNMR etched in HF and its zoomed view (e) reveal a high-fidelity perodic poling along the whole microring. Dark purple, inverted domains; light purple, uninverted domains. } 
\label{qpm}
\end{figure*}

First, since the optic axis of z-cut LN lies vertically, to employ its $d_{33}$ term, we design for phase-matching between the fundamental quasi-transverse-magnetic (TM) mode $a$ at 1560~nm and second-harmonic (SH) mode $b$ at 780~nm, as shown in Fig.~\ref{qpm}(a). The lower inset depicts the schematic cross-section of a partially etched z-cut LN microring with a radius of 70 $\mu$m. The fabrication of the z-cut air-cladded LN microrings is detailed in Ref.~\cite{Lu2019}. The lowest-order SH mode is favorable due to the lower scattering loss and larger modal overlap factor. The simulated profiles (amplitude of the vertical electric-field component) for the fundamental TM0 and SH TM0 modes are presented in the upper insets of Fig.~\ref{qpm}. Due to their large refractive index difference, the momentum conservation could only be satisfied via quasi-phase matching with a poling period of $\Lambda={\lambda_a}/{2(n_b-n_a)}=2.95~\mu$m as indicated by the double-headed black arrow in Fig.\ref{qpm}(a). Consequently, a final optimized $g/2\pi$ is calculated to be 1.78~MHz according to Eq.~(2) in Ref.~\cite{Lu2019d}. 

Second, for the implementation of quasi-phase matching, the poling electrodes with a tooth width of 750~nm are deposited on top of the etched LN microring, as shown in Fig.~\ref{qpm}(b). The tooth width is designed to be smaller than $\Lambda/2$ to allow for the inevitable lateral domain broadening and ensure a duty cycle of $\sim$ 50\%. The periodic ferroelectric domain inversion is then enabled by keeping the silicon substrate as the electrical ground while applying several 600~V, 250~ms pulses at an elevated temperature, as elaborated in Ref.~\cite{Lu2019d}. After removing the electrodes, piezoresponse force microscopy (PFM) is utilized as a non-destructive way to visualize the alternate domain inversion as presented in Fig.~\ref{qpm}(c), where the dark regions correspond to the inverted domains and a duty cycle close to 50\% is achieved. Moreover, the difference in etch rates between the poled and unpoled regions of z-cut LN in hydrofluoric acid (HF) allowed us to examine the poling quality along the whole microring under a scanning electron microscope (SEM). Figures~\ref{qpm}(d) and \ref{qpm}(e) are the false-color SEM images of a PPLNMR mock-up etched in HF, which imply a high-fidelity periodic poling. A period of $\sim$2.95~$\mu$m and a duty cycle of $\sim$50\% are confirmed. 

\section{Characterization of \lowercase{\boldmath{$g$}}}
The SHG measurement is implemented to verify the nonlinear coupling strength $g$ of the optimized PPLNMR device. With a pump field near the fundamental frequency, the system can be described by the Hamiltonian
\begin{eqnarray}
\hat{H}  = &&\omega_{a}{\hat{a}}^{\dagger}\hat{a} + \omega_b\hat{b}^{\dagger}\hat{b}  + g((\hat{a}^{\dagger})^2\hat{b} + \hat{a}^{2}\hat{b}^{\dagger}) \nonumber \\
&&+ i\epsilon_{p} (-\hat{a}e^{i\omega_{p}t}+\hat{a}^{\dagger}e^{-i\omega_{p}t}),
\end{eqnarray}
where $\epsilon_p = \sqrt{2\kappa_{a,1}P_{a,in}/\hbar\omega_p}$ is the input pump strength, $P_{a,in}$ is the on-chip pump power, and $\kappa_{a,1}$ denotes the external dissipation rate of the mode $a$.
At the steady state, the on-chip transmission of the pump laser and the output power of the SH signal are derived as 
\begin{eqnarray}
P_{a,out} = && \frac{\delta_a^2+(\kappa_{a,0}-\kappa_{a,1})^2}{\delta_a^2+\kappa_{a}^2}P_{a,in}, \label{transmission}\\
P_{b,out} = &&\frac{g^2P_{a,in}^2\omega_b}{\hbar{\omega_p}^2}\frac{2\kappa_{b,1}}{\delta_b^2+\kappa_{b}^2}(\frac{2\kappa_{a,1}}{\delta_a^2+\kappa_{a}^2})^2 \label{shgpower},
\end{eqnarray}
where $\kappa_{1}$, $\kappa_0$ and $\kappa$ are respectively the external, intrinsic and total dissipation rates of the cavity mode with $\kappa =\kappa_{1}+\kappa_0$. $\delta_{a}=\omega_{a}-\omega_{p}$ ($\delta_{b}=\omega_{b}-2\omega_{p}$) is the detuning for mode $a(b)$. The normalized SHG efficiency is given by $\eta_\mathrm{norm}=P_{b,out}/P_{a,in}^2$. Based on the measured $\kappa_{a(b)}$ and $\kappa_{a(b),1}$, the coupling rate $g$ could be fitted from the experimental SH response according to Eq.~(\ref{shgpower}).

\begin{figure*}
\includegraphics[width=1.0\textwidth]{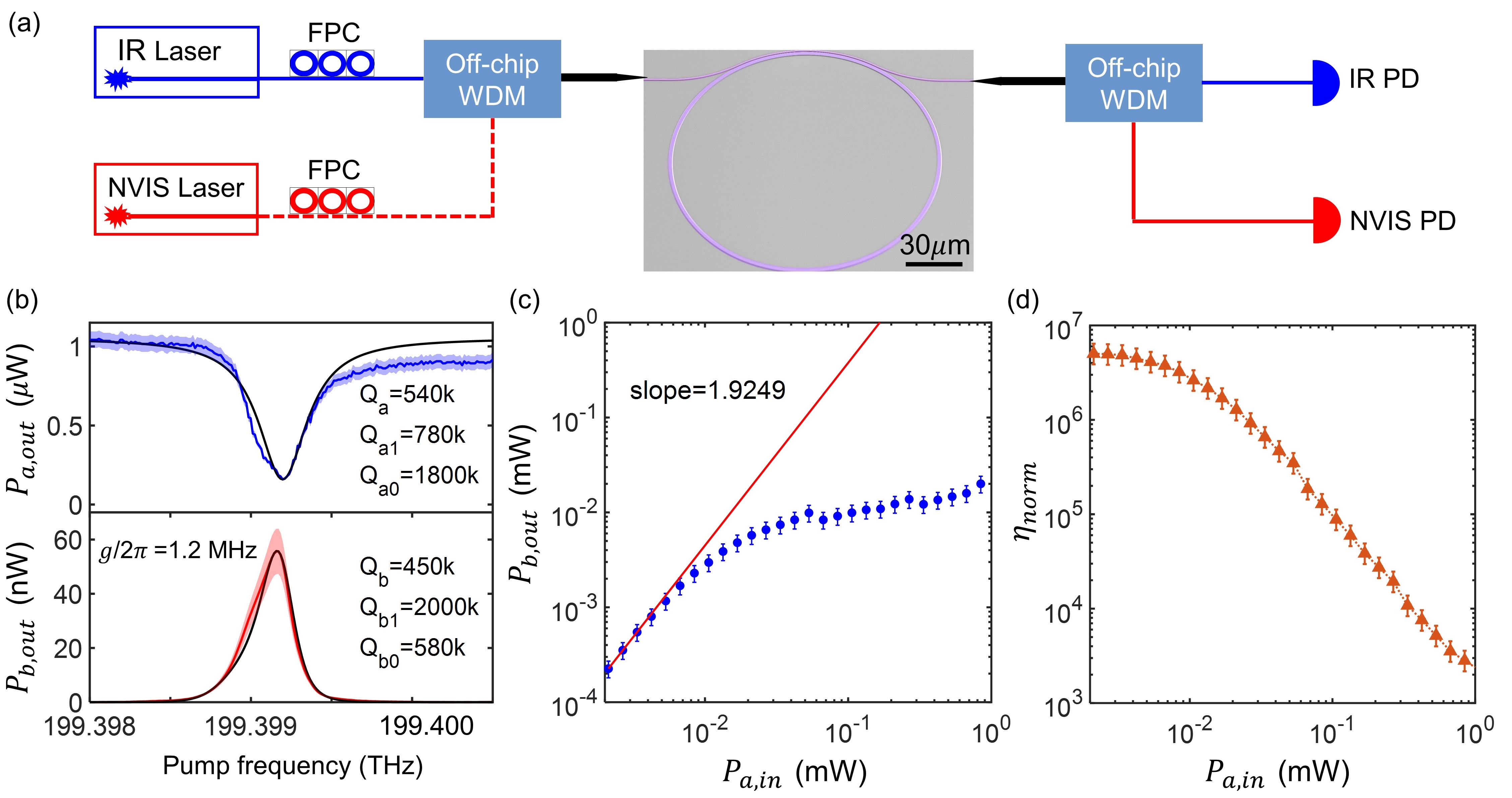}  
\caption{(a) Illustration of the experimental setup with a false-color SEM image of the PPLNMR device. IR: infrared, Nvis: near-visible, FPC: fiber polarization controller, WDM: Wavelength-division multiplexer, PD: photodetector. (b) Spectra of the pump resonance and corresponding SH response measured at an optimized temperature. (c) Quadratic relation is fitted at the low power regime while the deviation is observed as the pump power increases due to the intrinsic photorefractive effect. (d) Normalized conversion efficiency $\eta_\mathrm{norm}$ is plotted with the increasing pump power, indicating an optimized $\eta_\mathrm{norm}$ of $5,000,000\pm1,200,000$~\%/W at the low power regime whilst a decay in the high power regime.} 
\label{shg}
\end{figure*}

The experimental setup is illustrated in Fig.~\ref{shg}(a). 
The telecom and near-visible light sources are selectively turned on for the optical Q measurements, while only the telecom laser is swept for the SHG measurement. As shown in the upper panel of Fig.~\ref{shg}(b), the fundamental mode around 199~THz exhibits an intrinsic and loaded Q of $1.8\times 10^6$ and $5.4 \times10^5$, respectively. Likewise, the intrinsic and loaded Q factors for the SH mode are measured to be $5.8\times 10^5$ and $4.5\times 10^5$. Hence, the intrinsic dissipation rates of the fundamental and SH modes are respectively calculated to be $\kappa_{a,0}/2\pi =\omega_a/4\pi Q_{a0}= 55.4$~MHz and $\kappa_{b,0}/2\pi = 343.8$~MHz. Based on the calibrated insertion loss of 8.5 and 10.0~dB/facet for the respective fundamental and SH modes, Fig.~\ref{shg}(b) highlights the measured pump transmission and corresponding SH response at an optimal temperature, showing that a maximum on-chip SHG power of 55.6~nW is obtained with an on-chip pump power of 1.05~$\mu$W. The shaded region corresponds to an on-chip power variation induced by a coupling fluctuation of 5\% and 15\% for the respective fundamental and SH modes, and thereby an on-chip normalized SHG efficiency $\eta_\mathrm{norm}$ is estimated to be $5,000,000$~\%/W with an uncertainty of $1,200,000$~\%/W. 

The theoretical pump transmission and SHG output power based on Eqs.~(\ref{transmission}) and (\ref{shgpower}) are also plotted with the solid black lines in Fig.~\ref{shg}(b), where a nonlinear coupling strength $g/2\pi$ of 1.2~MHz is consequently fitted. The slight discrepancy between the measured $g/2\pi$ and the theoretically predicted value of 1.78~MHz is possibly due to nonuniformity inherent to nanofabrication at different azimuthal angle of the microring~\cite{luo2019} as well as random duty-cycle error, as implied in the SEM image of the selectively etched PPLNMR mock-up with hydrofluoric acid [Fig.~\ref{qpm}(d)]. Moreover, the power-dependence of the SHG power as well as the on-chip efficiency are plotted in Figs.~\ref{shg}(c) and \ref{shg}(d) respectively, where a linear-fitted slope of 1.92 justifies a quadratic dependence of SHG power on the pump power as predicted by Eq.~(\ref{shgpower}) and an $\eta_\mathrm{norm}$ of 5,000,000~\%/W is confirmed in the low power regime ($P_{a,in}<10~\mu$W). The deviation of the quadratic dependence and degradation of $\eta_\mathrm{norm}$ with the increasing pump power are probably attributed to the increasing frequency mismatch between the mode $a$ and $b$ induced by the accumulating photorefractive (PR) effect~\cite{preffectreview}. Such PR damage remains challenging for a broad class of thin film LN devices, including frequency converters and modulators~\cite{Lu2019d,mckenna2020cryogenic,1505236}, and will be investigated carefully in the future. As we are focusing on the nonlinearity at the single-quanta limit, the degraded device performance in the high power regime will not be a limiting factor for our device. The above experimental demonstration justifies the simultaneously optimized $g$ of 1.2~MHz and $\kappa_a/2\pi$ of 184~MHz via the record-high normalized SHG efficiency and highlights the potential of PPLNMR to play a key role in future quantum photonics applications.

\section{discussion and outlook}
\begin{figure}[htbp]
\includegraphics[width=0.5\textwidth]{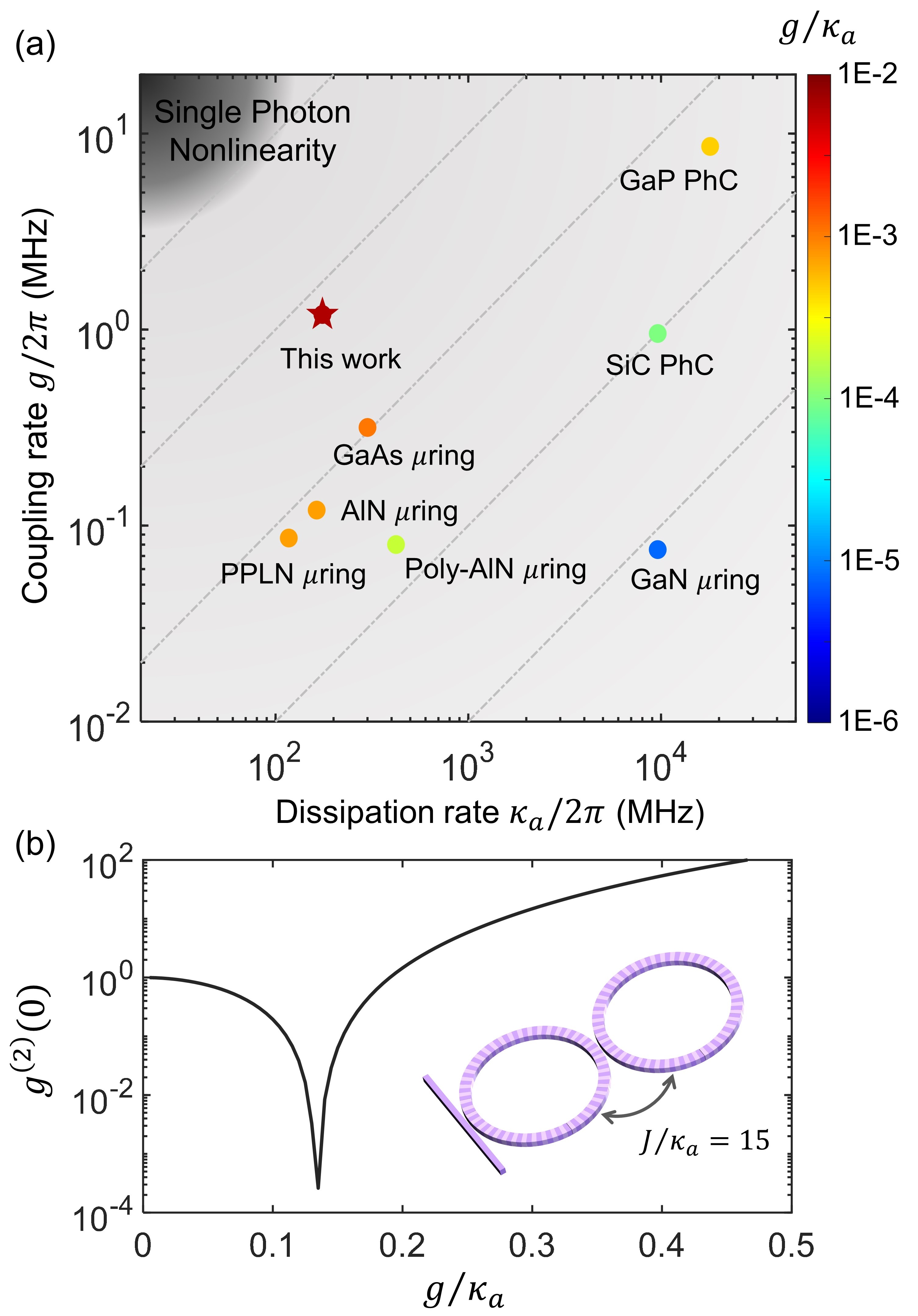}  
\caption{\label{diffplatforms}(a) The coupling rate $g$, dissipation rate $\kappa_a$ as well as the corresponding single-photon nonlinearity FOM demonstrated in various integrated $\chi^{(2)}$ photonics platforms, including AlN~\cite{Bruch2018a}, GaAs\footnote{Whose $g$ values are estimated from the representative data based on the assumptions that both fundamental and SH modes are critical-coupled and $\kappa_b=2\kappa_a$ when the $\kappa_b$ is not given.}~\cite{Chang2019a}, GaN$^{\text{a}}$~\cite{Roland2016}, poly-AlN~\cite{Guo2016e}, PPLN~\cite{Lu2019d} $\mu$rings, and GaP$^{\text{a}}$~\cite{Rivoire2010}, SiC$^{\text{a}}$~\cite{Yamada2014} PhCs. (b) Dependence of photon antibunching on $g/\kappa_a$. The inset presents a photonic molecule using two coupled PPLNMRs, allowing for a controllable linear coupling strength $J$ between the two fundamental modes in both PPLNMRs. For a $J/\kappa_a$ of 15, unconventional photon blockade occurs on condition that $g/\kappa_a$ reaches $10^{-1}$.} 
\end{figure}

The present PPLNMR device exhibits a state-of-the-art single photon nonlinearity FOM of 0.7$\times10^{-2}$ in comparison with that of the aluminum nitride (AlN)~\cite{Guo2016e,Surya2018,Bruch2018a}, gallium arsenide (GaAs)~\cite{Chang2019a}, gallium nitride (GaN)~\cite{Roland2016}, gallium phosphide (GaP)~\cite{Rivoire2010}, and silicon carbon (SiC)~\cite{Yamada2014} integrated $\chi^{(2)}$ cavities, as indicated in Fig.~\ref{diffplatforms}(a). By simply designing the external coupling to the under-coupled condition, the internal FOM ($g/\kappa_{a,0}$) of our device has already reached 0.02. The photonic crystal (PhC) provides another choice for high FOM by taking the advantage over the coupling rate $g$ due to its ultra-small mode volume ($\sim(\lambda/n)^3$), which is around three orders of magnitude smaller than that of the typical microrings. However, there are trade-offs in its relatively higher dissipation rate as the device is scaled down and also the difficulty in designing the simultaneous bandgap in fundamental and second-harmonic wavebands~\cite{Rivoire:s,Buckley:s,minkov2014optimizing,Lin:s}.

We note that although the strong coupling condition for photon blockade is not met by current device, a FOM smaller than unity still promises emitter-free quantum effect in photonic integrated circuits by employing the mechanism of unconventional photon blockade~\cite{PRLBlokade,PRAUBP,chi2UPB}. By designing interferometer in the Fock state space with an ancillary cavity mode, quantum states of deep sub-Poisson statistics have been demonstrated in quantum dot cavity QED~\cite{QDQED} and superconducting resonator~\cite{SuperCondBlockade}. Likewise, as shown in Fig.~\ref{diffplatforms}(b), by introducing another microring resonator and realizing a PPLNMR photonic molecule~\cite{zhang2019electronically}, single photons could be generated from a coherent laser input. Here, we set the linear coupling strength $J/\kappa_a=15$ between the two fundamental cavity modes in both coupled PPLNMRs. By further increasing coupling strength $g$ with a reduced microring radius of 40~$\mu$m while maintaining low dissipation rate $\kappa$ with a Q of 5 million through an optimized fabrication flow~\cite{zhang2017monolithic,ji2017ultra,zhang2019fabrication,liu2020high}, $g/2\pi$ of 2.35~MHz and $\kappa/2\pi$ of 19.2~MHz could be envisioned, which finally contributes to a FOM of 0.12 and thereby enables the efficient single photon anti-bunching, as indicated by the $g^{(2)}(0)$ dip approaching zero in Fig.~\ref{diffplatforms}(b).

\section{conclusion}
In conclusion, we have presented the optimization of $\chi^{(2)}$ photon-photon coupling strength towards single-photon nonlinearity in a PPLNMR. Utilizing its largest $\chi^{(2)}$ tensor element $d_{33}$, and implementing a high-fidelity radial poling period of 2.95~$\mu$m in an etched z-cut LN microring, a new-record normalized second-harmonic conversion efficiency of 5,000,000\%/W is demonstrated. Meanwhile, the corresponding single photon coupling rate $g/2\pi$ is estimated to be 1.2~MHz, which corresponds to a state-of-the-art FOM of $0.7\times 10^{-2}$ for single photon nonlinearity. With one order of magnitude improvement, we theoretically propose a PPLNMR photonic molecule device configuration that allows for the remarkable single-photon filtration via unconventional photon-blockade effect and paves the way for emitter-free, room-temperature quantum photonic applications, such as quantum light sources, photon-photon quantum gate, and quantum metrology.

\section{acknowledgments}
This work is supported by Department of Energy, Office of Basic Energy Sciences, Division of Materials Sciences and Engineering under Grant DE-SC0019406. HXT acknowledge partial funding supports from NSF (EFMA-1640959) and the Packard Foundation. The authors thank M. Rooks, Y. Sun for assistance in device fabrication.

\appendix

%\nocite{*}

\bibliography{References}% Produces the bibliography via BibTeX.

\end{document}